\documentclass[final,5p,times,twocolumn]{elsarticle}

\usepackage{amsmath}
\usepackage{amssymb}
\usepackage{graphicx}
\usepackage{pzccal}
\usepackage{color}
\usepackage{hyperref}

\definecolor{purple}{rgb}{0.5,0,0.5}
\definecolor{blue}{rgb}{0.0,0,0.9}

\biboptions{sort&compress}

\journal{Physics Letters B}

\begin{document}

\begin{frontmatter}

\title{Distribution amplitudes of light-quark mesons from lattice QCD}

\author[ANL]{Jorge Segovia}
\author[UA]{Lei Chang}
\author[ANL]{Ian C. Clo\"et}
\author[ANL]{Craig D. Roberts}
\author[JARA]{Sebastian M. Schmidt}
\author[Nanjing,Nanjing2,Nanjing3]{Hong-shi Zong}

\address[ANL]{Physics Division, Argonne National Laboratory, Argonne, Illinois 60439, USA}
\address[UA]{CSSM, School of Chemistry and Physics
University of Adelaide, Adelaide SA 5005, Australia}
\address[JARA]{Institute for Advanced Simulation, Forschungszentrum J\"ulich and JARA, D-52425 J\"ulich, Germany}
\address[Nanjing]{Department of Physics, Nanjing University, Nanjing 210093, China}
\address[Nanjing2]{State Key Laboratory of Theoretical Physics, Institute of Theoretical Physics, CAS, Beijing 100190, China}
\address[Nanjing3]{Joint Center for Particle, Nuclear Physics and Cosmology, Nanjing 210093, China}

\date{19 October 2013}

\begin{abstract}
We exploit a method introduced recently to determine parton distribution amplitudes (PDAs) from minimal information in order to obtain light-quark pseudoscalar and vector meson PDAs from the limited number of moments produced by numerical simulations of lattice-regularised QCD.  Within errors, the PDAs of pseudoscalar and vector mesons constituted from the same valence quarks are identical; they are concave functions, whose dilation expresses the strength of dynamical chiral symmetry breaking; and SU$(3)$-flavour symmetry is broken nonperturbatively at the level of 10\%.  Notably, the appearance of precision in the lattice moments is misleading.  The moments also exhibit material dependence on lattice volume, especially for the pion.  Improvements need therefore be made before an accurate, unified picture of the light-front structure of light-quark pseudoscalar and vector mesons is revealed.
\end{abstract}

\begin{keyword}
quantum chromodynamics \sep dynamical chiral symmetry breaking \sep Dyson-Schwinger equations \sep lattice-regularised QCD \sep light-front quantum field theory \sep light mesons \sep parton distribution amplitudes \sep strange quarks

\smallskip

\emph{Preprint no}. ADP-13-21/T841



\end{keyword}

\end{frontmatter}

\noindent\textbf{1.$\;$Introduction}.
%
%
A valence parton distribution amplitude (PDA) is a light-front wave-function of an interacting quantum system.  It provides a connection between dynamical properties of the underlying relativistic quantum field theory and notions familiar from nonrelativistic quantum mechanics.  In particular, although particle number conservation is generally lost in relativistic quantum field theory, $\varphi(x)$ has a probability interpretation.  It can therefore translate features that arise purely through the infinitely-many-body nature of relativistic quantum field theory into images whose interpretation seems more straightforward \cite{Keister:1991sb,Coester:1992cg,Brodsky:1997de,Chang:2013pq,Cloet:2013tta,Chang:2013epa}.  For a meson, the argument of the PDA, $x$, expresses the light-front fraction of the bound-state total-momentum carried by the meson's valence quark, which is equivalent to the momentum fraction carried by the valence-quark in the infinite-momentum frame; and momentum conservation entails that the valence antiquark carries the fraction $\bar x=(1-x)$.

In the theory of strong interactions, the cross-sections for many \emph{hard} exclusive hadronic reactions can be expressed accurately in terms of the PDAs of the hadrons involved \cite{Lepage:1979zb,Farrar:1979aw,Lepage:1980fj,Efremov:1979qk,Beneke:1999br,Beneke:2000ry,Beneke:2001ev}.  For example, in the case of the electromagnetic form factor of light pseudoscalar mesons  \cite{Efremov:1979qk,Lepage:1979zb,Farrar:1979aw,Lepage:1980fj}:
\begin{align}
\label{pionUV}
\exists Q_0 > & \Lambda_{\rm QCD} \; |\;   Q^2 F_P(Q^2) \stackrel{Q^2 > Q_0^2}{\approx} 16 \pi \alpha_s(Q^2)  f_P^2 \mathpzc{w}_\varphi^2,  \\
& \label{wphi}
\mathpzc{w}_\varphi = \frac{1}{3} \int_0^1 dx\, \frac{1}{x}\, \varphi^P(x)\,,
\end{align}
where $\alpha_s(Q^2)$ is the strong running coupling, $f_P$ is the meson's leptonic decay constant and $\varphi^P(x)$ is its PDA.  Such formulae are exact.  However, the PDAs are not determined by the analysis framework; and the value of $Q_0$ is not predicted.  (N.B.\ Dynamical generation of the mass-scale $\Lambda_{\rm QCD} \sim 0.2\,$GeV in QCD spoils the conformal invariance of classical massless QCD \cite{Collins:1976yq,Nielsen:1977sy,tarrach} and is very likely connected intimately with the emergent phenomena of confinement and dynamical chiral symmetry breaking (DCSB) \cite{Chang:2013epa}.)

One may alternatively describe exclusive reactions in terms of Poincar\'e-covariant hadron bound-state amplitudes (BSAs), obtained from Bethe-Salpeter or Faddeev equations.  This approach has been used widely; e.g., \cite{Jarecke:2002xd,Maris:2003vk,Arrington:2006zm,Holt:2010vj,Eichmann:2011ej,%
Aznauryan:2012ba,Chang:2011vu,Bashir:2012fs,Cloet:2013jya}: in explaining reactions used to measure elastic and transition electromagnetic form factors, another class involving strong decays of hadrons, and yet another group relating to the semileptonic and nonleptonic weak decays of heavy mesons.  The BSAs are predictions of the framework and the associated computational scheme is applicable on the entire domain of accessible momentum transfers.  However, truncations must be employed in formulating the problem; and issues related to the construction of veracious truncation schemes are canvassed, e.g., in Ref.\,\cite{Chang:2011vu,Bashir:2012fs,Cloet:2013jya}.

The two approaches are joined by the fact that the PDAs, which are essentially nonperturbative, may be obtained as light-front projections of the hadron BSAs.  Recent progress has established this connection as a practical reality and thereby produced a particularly effective synergy \cite{Chang:2013pq}, which is highlighted by a prediction for a pion's elastic electromagnetic form factor \cite{Chang:2013nia}.  These advances can be summarised succinctly: it is now possible to compute bound-state PDAs from Poincar\'e-covariant hadron BSAs and thereby place useful and empirically verifiable constraints on both.

Two significant features emerged in developing the connection between BSAs and PDAs.  The first is an appreciation that the so-called asymptotic PDA; i.e.,
\begin{equation}
\varphi^{\rm asy}(x) = 6 x \bar x\,, \label{phiasy}
\end{equation}
provides an unacceptable description of meson internal structure at all scales which are either currently accessible or foreseeable in experiments \cite{Cloet:2013jya,Chang:2013pq,Chang:2013nia,Cloet:2013tta}.  This should not be surprising because evolution with energy scale in QCD is \emph{logarithmic}.

The second important point is that at all energy scales, the leading twist PDAs for light-quark mesons are concave functions.  This last is a powerful observation because it eliminates the possibility of ``humped'' distributions \cite{Chernyak:1983ej} and enables one to obtain a pointwise accurate approximation to a meson's valence-quark PDA from a very limited number of moments \cite{Cloet:2013tta}, which is all that is available, e.g., from numerical simulations of lattice-regularised QCD \cite{Braun:2006dg,Braun:2007zr,Arthur:2010xf}.

An effort has recently begun, focused on computation of meson PDAs directly from BSAs obtained using QCD's Dyson-Schwinger equations (DSEs) \cite{Chang:2011vu,Bashir:2012fs,Cloet:2013jya}.  With the well constrained kernels for bound-state equations that are now available \cite{Chang:2009zb,Chang:2011ei,Qin:2011dd,Qin:2011xq}, these studies have a direct connection to QCD; and hence comparison of their results with experiment will serve as meaningful tests of this theory, as were previous computations of parton distribution functions  \cite{Hecht:2000xa,Holt:2010vj,Aicher:2010cb,Nguyen:2011jy}.

Independently, it is worth capitalising on the second observation reported above; namely, to use extant results from numerical simulations of QCD in order to obtain insights into the pointwise behaviour of meson PDAs, as has already been tried for parton distribution functions (PDFs) \cite{Detmold:2001dv,Detmold:2003tm,Detmold:2003rq}.  The results should be valuable in the analysis and planning of contemporary and future experiments.  They will also serve as a benchmark by which to gauge outcomes of attempts at the computation of meson PDAs, including the DSE studies already mentioned but also results from QCD sum-rules (e.g., Refs.\,\cite{Mikhailov:1986be,Ball:1996tb,Bakulev:2001pa,Khodjamirian:2004ga,Ball:2006fz,%
Agaev:2010aq,Bakulev:2012nh}) and models (e.g., Refs.\,\cite{Frederico:1994dx,Brodsky:2006uqa,Dorokhov:2006xw,Choi:2007yu,Ahmady:2012dy}).
\smallskip

\noindent\textbf{2.$\;$Computing PDAs from moments}.
One should properly denote a meson PDA by $\varphi(x;\tau)$.  It is a function of two arguments: $x$, the parton light-front momentum fraction; and $\tau=1/\zeta$, where $\zeta$ is the momentum-scale that characterises the exclusive process in which the meson is involved.  On the domain within which QCD perturbation theory is valid, the equation describing the $\tau$-evolution of $\varphi(x;\tau)$ is known and has the solution  \cite{Efremov:1979qk,Lepage:1980fj}
\begin{eqnarray}
\label{PDAG3on2}
\varphi(x;\tau) &=& \varphi^{\rm asy}(x)
\bigg[ 1 + \!\! \sum_{j=1,2,\ldots}^{\infty} \!\! \!\! a_j^{3/2}(\tau) \,C_j^{(3/2)}(x -\bar x) \bigg],\;\;
\end{eqnarray}
where $\{C_j^{(3/2)},j=1,2,\ldots,\infty\}$ are Gegenbauer polynomials of order $\alpha=3/2$ and the expansion coefficients $\{a_j^{3/2},j=1,2,\ldots,\infty\}$ evolve logarithmically with $\tau$, vanishing as $\tau\to 0$.  This result expresses the fact that in the neighbourhood $\tau \Lambda_{\rm QCD} \simeq 0$, QCD is invariant under the collinear conformal group
SL$(2;\mathbb{R})$
\cite{Brodsky:1980ny,Braun:2003rp}.  Gegenbauer-$\alpha=3/2$ polynomials are the irreducible representations of this group.  A correspondence with the spherical harmonics expansion of the wave functions for $O(3)$-invariant systems in quantum mechanics is plain.

Nonperturbative methods in QCD typically provide access to moments of the PDA; viz., the quantities
\begin{equation}
\langle (x -\bar x)^m \rangle_\varphi^\tau = \int_0^1\!dx\,(x -\bar x)^m\, \varphi(x;\tau)\,.
\end{equation}
Until recently it was commonly assumed that at any length-scale $\tau$, an accurate approximation to $\varphi(x;\tau)$ is obtained by using just the first few terms of the expansion in Eq.\,\eqref{PDAG3on2}; and hence that the best use of a limited number of moments was to determine the first few Gegenbauer coefficients, $a_j^{3/2}(\tau)$, in Eq.\,\eqref{PDAG3on2}.  We will call this \emph{Assumption~A}.  It leads to models for $\varphi(x)$ whose pointwise behaviour is not concave on $x\in[0,1]$; e.g., to ``humped'' distributions \cite{Chernyak:1983ej}.  Following Ref.\,\cite{Chang:2013pq}, one may readily establish that a double-humped form for $\varphi(x)$ lies within the class of distributions produced by a meson BSA which may be characterised as vanishing at zero relative momentum, instead of peaking thereat.  No ground-state pseudoscalar or vector meson solution exhibits such behaviour \cite{Maris:1997tm,Maris:1999nt,Qin:2011xq}.

\emph{Assumption~A} is certainly valid on $\tau \Lambda_{\rm QCD} \simeq 0$.  However, it is grossly incorrect at any energy scale accessible in contemporary or foreseeable experiments.  This was highlighted in Ref.\,\cite{Cloet:2013tta} and in Sec.\,5.3 of Ref.\,\cite{Cloet:2013jya}.  The latter used the fact \cite{Georgi:1951sr,Gross:1974cs,Politzer:1974fr} that $\varphi^{\rm asy}(x)$ can only be a good approximation to a meson's PDA when it is accurate to write $u_{\rm v}(x) \approx \delta(x)$, where $u_{\rm v}(x)$ is the meson's valence-quark PDF, and showed that this is not valid even at energy scales characteristic of the large hadron collider (LHC).  Hence, realistic meson PDAs are necessarily much broader than $\varphi^{\rm asy}(x)$.  It follows that an insistence on using just a few terms in Eq.\,\eqref{PDAG3on2} to represent a hadron's PDA must lead to unphysical oscillations; i.e., humps, just as any attempt to represent a box-like curve via a Fourier series will inevitably lead to slow convergence and spurious oscillations.

An alternative to \emph{Assumption A}, advocated and explained in Refs.\,\cite{Chang:2013pq,Cloet:2013tta,Chang:2013epa,Cloet:2013jya}, is to accept that at all accessible scales, the pointwise profile of PDAs is determined by nonperturbative dynamics; and hence PDAs should be reconstructed from moments by using Gegenbauer polynomials of order $\alpha$, with this order -- the value of $\alpha$ -- determined by the moments themselves, not fixed beforehand.  In illustrating this procedure, Ref.\,\cite{Chang:2013pq} considered DSE results for the pion's BSA, wrote
\begin{equation}
\label{PDAGalpha}
\varphi(x;\tau) = N_\alpha\, [x \bar{x}]^{\alpha_-}
\bigg[ 1 + \sum_{j=2,4,\ldots}^{j_s} \!\!\!\! a_j^{\alpha}(\tau)\, C_j^{(\alpha)}(x-\bar x) \bigg],\quad
\end{equation}
where $\alpha_-=\alpha-1/2$ and $N_\alpha = 1/B(\alpha+1/2,\alpha+1/2)$
and obtained a converged, concave result for the PDA with $j_s=2$.  (N.B.\ In the case of mesons in a multiplet that contains an eigenstate of charge-conjugation, $\varphi(x) = \varphi(\bar x)$; and hence only even terms contribute to the sum in Eq.\,\eqref{PDAG3on2}.)  Naturally, once obtained in this way, one may project $\varphi(x;\tau)$ onto the form in Eq.\,\eqref{PDAG3on2}; viz., for $j=1,2,\ldots\,$,
\begin{equation}
\label{projection}
a_j^{3/2}(\tau) = \frac{2}{3}\ \frac{2\,j+3}{(j+2)\,(j+1)}\int_0^1 \!dx\, C_j^{(3/2)}(x-\bar x)\,\varphi(x;\tau),
\end{equation}
therewith obtaining all coefficients necessary to represent any computed distribution in the conformal form without ambiguity or difficulty.  In this form, too, one may determine the distribution at any $\tau^\prime < \tau$ using the ERBL evolution equations for the coefficients $\{a_j^{3/2}(\tau),i=1,2,\ldots\}$ \cite{Efremov:1979qk,Lepage:1980fj}.

\begin{table}[t]
\caption{Meson PDA moments obtained using numerical simulations of lattice-regularised QCD with $N_f = 2+1$ domain-wall fermions and nonperturbative renormalisation of lattice operators \protect\cite{Arthur:2010xf}: linear extrapolation to physical pion mass, $\overline{\rm MS}$-scheme at $\zeta=2\,$GeV, two lattice volumes.   The first error is statistical, the second represents an estimate of systematic errors, including those from the $s$-quark mass, discretisation and renormalisation. \label{latticemoments}}
\begin{center}
\begin{tabular}{ll|ll}\hline
\rule{0em}{2.3ex}
meson & $\langle (x-\bar x)^n\rangle$ & $16^3 \times 32$ & $24^3 \times 64$ \\\hline
$\pi$ & n$=$2 & 0.25(1)(2) & 0.28(1)(2) \\
$\rho_\parallel$ & n$=$2 & 0.25(2)(2) & 0.27(1)(2)\\
$\phi$ & n$=$2 &  0.25(2)(2) & 0.25(2)(1) \\
$K$ & n$=$1 & 0.035(2)(2) & 0.036(1)(2) \\
$K^\ast_\parallel$ & n$=$1 & 0.037(1)(2) & 0.043(2)(3) \\
$K$ & n$=$2 & 0.25(1)(2) & 0.26(1)(2) \\
$K^\ast_\parallel$ & n$=$2 & 0.25(1)(2) & 0.25(2)(2) \\\hline
\end{tabular}
\end{center}
\end{table}

In connection with the challenge of reconstructing a distribution from moments, consider that since discretised spacetime does not possess the full rotational symmetries of the Euclidean continuum, then, with current algorithms, at most two nontrivial moments of $\varphi(x)$ can be computed using numerical simulations of lattice-regularised QCD.  In the case of mesons in a multiplet that contains an eigenstate of charge-conjugation one has $\langle x -\bar x \rangle \equiv 0$, which means that, on average, the valence-quark and -antiquark share equally in the light-front momentum of the bound-state; and hence only one nontrivial moment is accessible.  Herein we propose to follow Ref.\,\cite{Cloet:2013tta} and use this limited information to reconstruct PDAs from lattice-QCD moments using an analogue of Eq.\,\eqref{PDAGalpha} that is also valid for mesons comprised from valence-quarks with nondegenerate masses:
\begin{equation}
\varphi(x) = x^\alpha \, (1-x)^\beta / B(\alpha,\beta).
\label{eqFitphi}
\end{equation}
The moments listed in Table~\ref{latticemoments} are sufficient to determine $\alpha$, $\beta$ in all instances; and, as mentioned above, if one wishes to evolve the distribution obtained to another momentum scale, $\tau^\prime<\tau$, then this may be achieved by projecting Eq.\,\eqref{eqFitphi} onto the form in Eq.\,\eqref{PDAG3on2} using Eq.\,\eqref{projection}, and subsequently employing the ERBL evolution equations \cite{Efremov:1979qk,Lepage:1980fj}.

\smallskip

\noindent\textbf{3.$\;$Light pseudoscalar and vector mesons with equal-mass valence-quarks}.
Consider now that a vector meson has two PDAs, one associated with light-front longitudinal polarisation, $\varphi^V_\parallel$, and the other with light-front transverse polarisation, $\varphi^V_\perp$.  Simulations of lattice-QCD performed thus far have produced $\tau_2=1/\zeta_2$, $\zeta_2=2\,$GeV, moments of $\varphi^V_\parallel$ and $\varphi^V_\perp$ which are equal within errors \cite{Braun:2007zr}.  Similarly, it is apparent in Table~\ref{latticemoments} that contemporary lattice-QCD cannot distinguish between $\varphi^V_\parallel$ and $\varphi^P$, where the latter is the PDA associated with the vector meson's pseudoscalar analogue.  We expect, however, that in reality these PDAs are different.  Indeed,  since a vector meson's electric radius is greater than its magnetic radius, and the latter, in turn, is greater than the charge radius of the pseudoscalar meson analogue \cite{Bhagwat:2006pu,Roberts:2011wy}, we anticipate the following ordering at accessible energy scales:
\begin{equation}
\varphi^V_\parallel\; \mbox{narrower-than} \; \varphi^V_\perp \;  \mbox{narrower-than} \; \varphi^P\,,
\label{narrower}
\end{equation}
where ``narrower'' means pointwise closer to $\varphi_\pi^{\rm asy}(x)$.  This expectation requires confirmation via explicit calculations within the same DSE framework that delivered the stated ordering of radii.

The need for such a study is highlighted by the following observations.
The pattern of Eq.\,\eqref{narrower} is seen in Refs.\,\cite{Ball:1996tb,Choi:2007yu}, which report $\varphi^V_\parallel$ a little narrower than $\varphi^V_\perp$, and both narrower than $\varphi^P$.
In contrast, combining Refs.\,\cite{Brodsky:2006uqa,Ahmady:2012dy} one finds $\varphi^V_\parallel(x) \approx \varphi^P(x)$ but $\varphi^P$ much narrower than $\varphi^V_\perp$, whereas
Ref.\,\cite{Dorokhov:2006xw} produces $\varphi^V_\parallel(x) \approx \varphi^P(x)$ but $\varphi^V_\perp$ much narrower than $\varphi^V_\parallel$.
%

The inconsistency just described is plainly unsatisfactory.  So, absent a well-constrained DSE study, herein we simply work with the contemporary lattice-QCD result: $\varphi^V_\parallel \approx \varphi^V_\perp \approx \varphi^P=: \varphi_{du}$, and report PDAs obtained from the pseudoscalar moments in Table~\ref{latticemoments}.  Using Eq.\,\eqref{eqFitphi}, the two rightmost columns of this Table yield:
\begin{eqnarray}
\label{phi16} 16^3 \times 32{:} && \alpha_{du} = \beta_{du} = 0.50^{+0.20}_{-0.16}\,, \\
\label{phi24} 24^3 \times 64{:} && \alpha_{du} = \beta_{du} = 0.29^{+0.15}_{-0.13} \,.
\end{eqnarray}

The PDAs in Eqs.\,\eqref{phi16}, \eqref{phi24} precisely reproduce the values of the moments in Table~\ref{latticemoments} and predict the quantities listed in Table~\ref{predictions}.
%

\begin{table}[t]
\caption{Selection of computed quantities associated with the meson PDAs in Eqs.\,\eqref{phi16}, \eqref{phi24}, \eqref{phiK16}, \eqref{phiK24}.  $x_{\rm max}$ is the location of the PDA's maximum, which lies at $x=\frac{1}{2}$ for the $du$ case ($\varphi^{\rm asy}(x_{\rm max})=1.5$), and $\mathpzc{w}$ is defined in Eq.\,\eqref{wphi}.   The fact that the $n=3,4$ moments have values which are $\gtrsim 60$\% of their kindred lower moments highlights the statements made in connection with Eq.\,\eqref{PDAGalpha}; i.e., that any attempt to reconstruct the PDA using Eq.\,\eqref{PDAG3on2} must converge very slowly. \label{predictions}}
\begin{center}
\begin{tabular}{l|ll}\hline
\rule{0em}{2.3ex} & $16^3\times 32$ & $24^3\times 64$ \\\hline
$\rule{0em}{2ex} \varphi_{du}(x_{\rm max})$ & $1.27^{+0.09}_{-0.08}$&$1.16^{+0.08}_{-0.07}$ \\[1.2ex]
$\rule{0em}{2ex}\varphi_{su}(x_{\rm max})$ & $1.28^{+0.09}_{-0.08}$
& $1.24^{+0.09}_{-0.08}$\rule{0ex}{1.8ex}\\[1.2ex]
$\rule{0em}{2ex}\langle (x-\bar x)^3\rangle_{su}$ & $0.019$ & $0.020$ \\
$\rule{0em}{2ex}\langle (x-\bar x)^4\rangle_{su}$ & $0.13 \pm 0.02$ & $0.13 \pm 0.02$\\[1.2ex]
$\rule{0em}{2ex}\langle (x-\bar x)^4\rangle_{du}$ & $0.125^{+0.019}_{-0.018}$ & $0.15 \pm 0.02$\\[1.2ex]
$\mathpzc{w}_{du}$ & $1.33^{+0.32}_{-0.19}$ & $1.83^{+1.00}_{-0.41}$ \\[1.2ex]
$\mathpzc{w}_{su}$ & $1.20^{+0.26}_{-0.16}$ & $1.29^{+0.33}_{-0.19}$ \\[1.2ex]
\hline
\end{tabular}
\end{center}
\end{table}

It is worth remarking here that there are two extremes for the PDA: $\varphi_{du}=\varphi^{\rm point}\,=\,$constant, which describes a point-particle; and $\varphi_{du}=\,\varphi^{\rm asy}$, which is the result in conformal QCD.  This means that the second moment is bounded as follows:
\begin{equation}
\frac{1}{2}=\langle (x-\bar x)^2 \rangle_{\varphi^{\rm asy}}
\leq \langle (x-\bar x)^2 \rangle_{\varphi} \leq
\langle (x-\bar x)^2 \rangle_{\varphi^{\rm point}} = \frac{1}{3}\,.\quad
\end{equation}
Therefore, instead of using an absolute scale, the accuracy of and deviations between the moments in Table~\ref{latticemoments} should be measured against these bounds.  Consequently, the difference between the central values of the $\pi$\,--\,$n=2$ entries in the second row of the Table corresponds to a mismatch of 23\%.  This explains the marked differences between Eqs.\,\eqref{phi16} and \eqref{phi24}.  The analogous bounds on the fourth moment are given by $3/35 (=0.086) < \langle (x-\bar x)^4 \rangle < 1/5$ and should be borne in mind when reflecting upon Table~\ref{predictions}.

\begin{figure}[t]
\centerline{\includegraphics[width=0.9\linewidth]{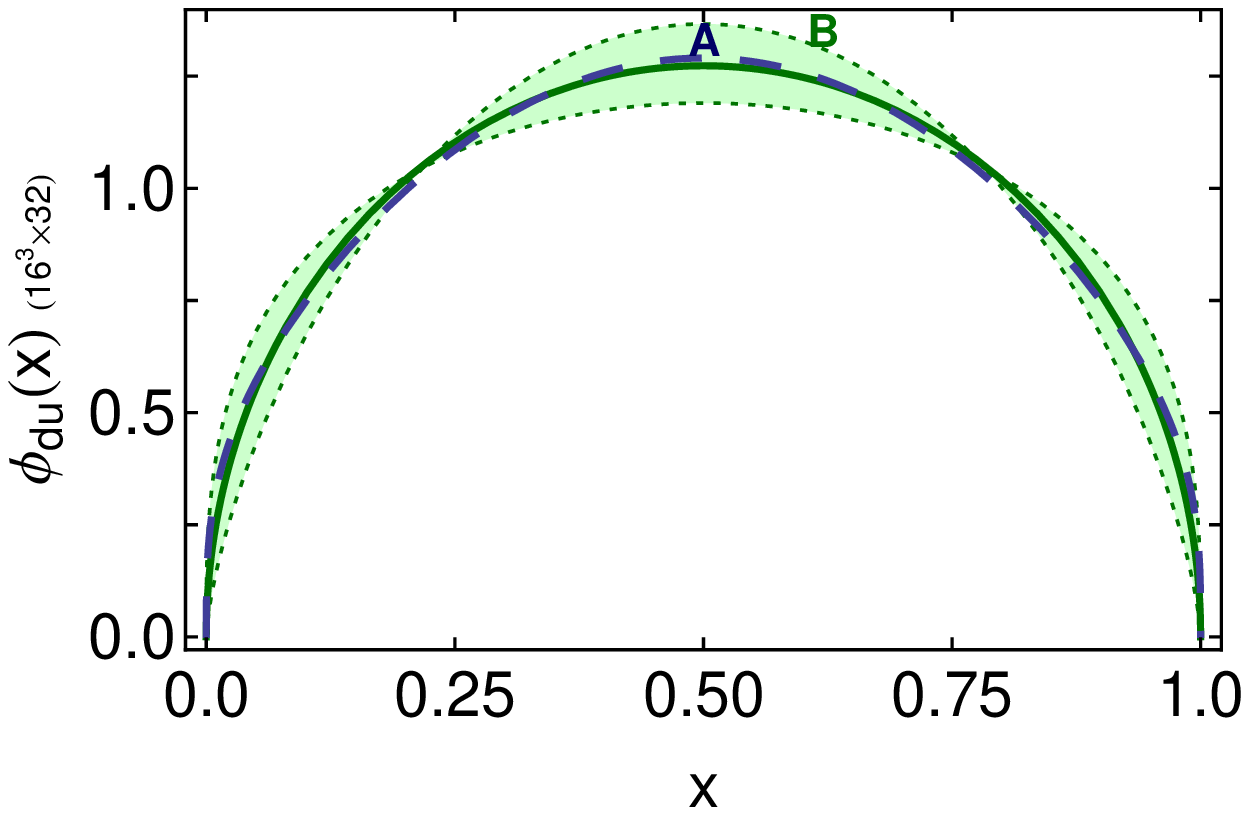}}
\centerline{\includegraphics[width=0.9\linewidth]{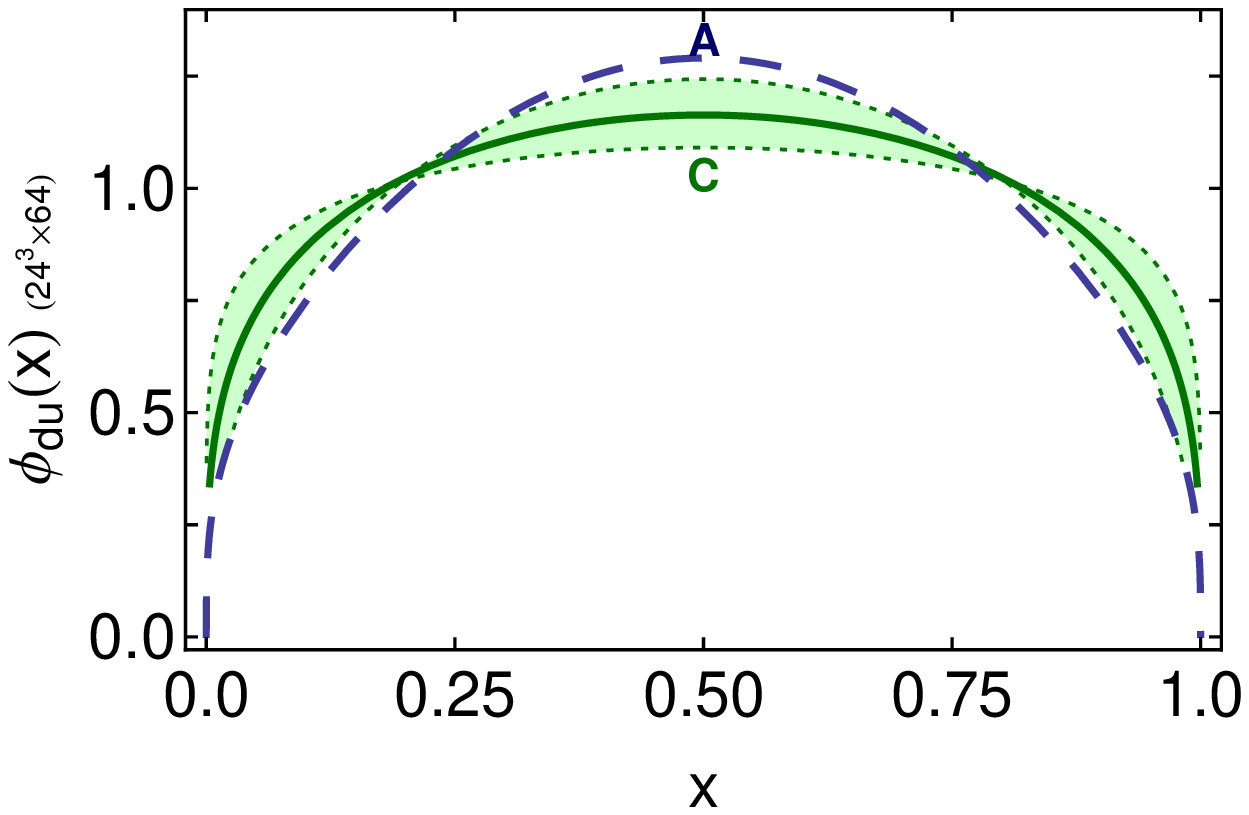}}
\caption{PDA for pseudoscalar and vector mesons constituted from equal mass valence-quarks, reconstructed using Eq.\,\eqref{eqFitphi}.
\emph{Upper panel} -- solid curve and associated error band (shaded region labelled ``B''): Eq.\,\eqref{phi16}, obtained from the $16^3\times 32$-lattice moments in Table~\ref{latticemoments};
and \emph{lower panel} -- solid curve and associated error band (shaded region labelled ``C''): Eq.\,\eqref{phi24}, obtained from the $24^3\times 64$-lattice moments in Table~\ref{latticemoments}.  The dashed curve ``A'' in both panels is the DSE prediction in Eq.\,\eqref{resphipi2DB}.
\label{figNewLattice}}
\end{figure}

Region~B in the upper panel of Fig.\,\ref{figNewLattice} displays the result in Eq.\,\eqref{phi16}, with the interior solid curve marking $\alpha_{du}=0.50$; and Region~C in the lower panel of Fig.\,\ref{figNewLattice} depicts the result in Eq.\,\eqref{phi24}, with the interior solid curve marking $\alpha_{du}=0.29$.  The dashed curve labelled ``A'' in both panels is the DSE prediction for the chiral-limit pion:
\begin{equation}
\label{resphipi2DB}
\varphi_\pi(x;\tau_2) = 1.81 [x (1-x)]^{\mathpzc a} \, [1 + \tilde a_2 C_2^{\mathpzc{a}+1/2}(2 x - 1)]\,,
\end{equation}
$\mathpzc{a} = 0.31$, $\tilde a_2=-0.12$, which was obtained elsewhere \cite{Chang:2013pq} using the most sophisticated symmetry-preserving kernels for the gap and Bethe-Salpeter equations that are currently available \cite{Chang:2011ei}.  These kernels incorporate essentially nonperturbative effects associated with DCSB, which are omitted in the leading-order (rainbow-ladder) truncation and any stepwise improvement thereof \cite{Chang:2009zb}.  They have exposed a key role played by the dressed-quark anomalous chromomagnetic moment \cite{Chang:2010hb} in determining observable quantities; e.g., clarifying a causal connection between DCSB and the splitting between vector and axial-vector mesons \cite{Chang:2011ei}.
%
If one chooses to approximate Eq.\,\eqref{resphipi2DB} via Eq.\,\eqref{eqFitphi}, then it corresponds to $\alpha=\beta=0.50$, which is also the value associated with the models described in Refs.\,\cite{Mikhailov:1986be,Brodsky:2006uqa}.

Overlaying the two panels of Fig.\,\ref{figNewLattice}, one finds that the PDAs obtained from the two different lattice spacings have overlapping error bands.  Notwithstanding this, the differences are material, something which may be illustrated by considering the ``$1/x$'' moment of the PDAs that, according to Eqs.\,\eqref{pionUV}, \eqref{wphi}, sets the large-$Q^2$ magnitude in the perturbative QCD formulae for a pseudoscalar meson's elastic form factor.  These moments are presented in Table~\ref{predictions}.
The DSE prediction is $\mathpzc{w}_{du}=(1/3)\langle x^{-1}\rangle=1.53$, a result compatible with that obtained using the PDAs of Refs.\,\cite{Mikhailov:1986be,Brodsky:2006uqa}; and a QCD sum rules analysis produces \cite{Bakulev:2001pa} $\mathpzc{w}_{du}=1.1\pm0.1$.  These continuum-QCD results are compatible with experiment.  It appears, therefore, that the $24^3 \times 64$ lattice configurations produce a form of $\varphi_{du}(x)$ that is too broad.

\begin{figure}[t]
\centerline{\includegraphics[width=0.9\linewidth]{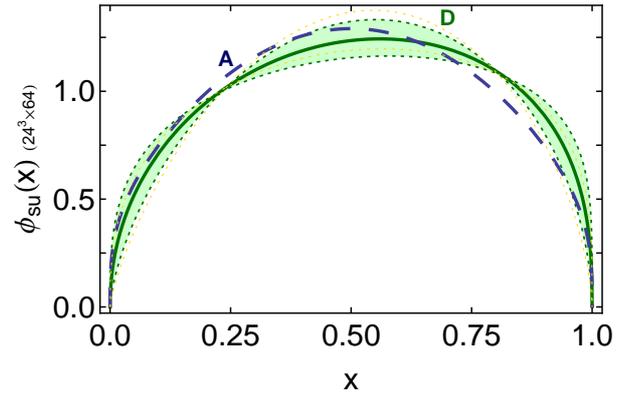}}
\caption{Solid curve and associated error band (shaded region labelled ``D''): PDA in Eq.\,\eqref{phiK24}, describing $s \bar u $ pseudoscalar and vector mesons, reconstructed using Eq.\,\eqref{eqFitphi} and obtained from the $24^3\times 64$-lattice configurations.  The result obtained from the $16^3\times 32$-lattice moments in Table~\ref{latticemoments} is not materially different.  The dashed curve ``A'' is the DSE prediction for the pion's PDA in Eq.\,\eqref{resphipi2DB}.
\label{figNewLatticeK}}
\end{figure}

\begin{figure}[t]
\centerline{\includegraphics[width=0.9\linewidth]{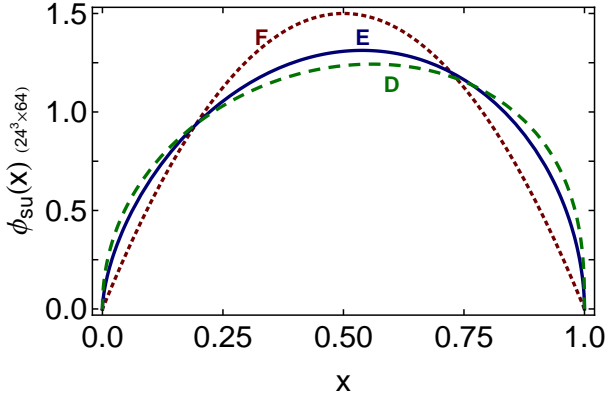}}
\caption{Solid curve (labelled ``E'') is ERBL evolution to $\zeta_{10}=10\,$GeV of kaon PDA defined by Eq.\,\eqref{phiK24} (dashed curve, labelled ``D'' to match the same PDA in Fig.\,\ref{figNewLatticeK}).  Dotted curve (labelled ``F'') is $\varphi^{\rm asy}(x)$ in Eq.\,\eqref{phiasy}.
\label{figEvolvedK}}
\end{figure}

The preceding analysis emphasises anew that information is gained using the procedure advocated in Refs.\,\cite{Chang:2013pq,Cloet:2013jya,Chang:2013epa,Cloet:2013tta} but not lost.  It has enabled an informed analysis of the lattice results, providing context and highlighting possible shortcomings.

\smallskip

\noindent\textbf{4.$\;$\mbox{\boldmath $s \bar u$} pseudoscalar and vector mesons}.
When reconstructing a PDA for $s \bar u$ mesons, we choose to focus on the pseudoscalar meson moments in Table~\ref{latticemoments} because they show the least sensitivity to lattice volume and possess the smallest errors.  Using the procedure described in association with Eq.\,\eqref{PDAGalpha}, they yield
\begin{eqnarray}
\label{phiK16} 16^3 \times 32{:} && \alpha_{su} = 0.56^{+0.21}_{-0.18}\,, \;
\beta_{su} = 0.45^{+0.19}_{-0.16}\,, \\
\label{phiK24} 24^3 \times 64{:} && \alpha_{su} = 0.48^{+0.19}_{-0.16}\,,\;
\beta_{su} = 0.38^{+0.17}_{-0.15} \,.
\end{eqnarray}
These PDAs precisely reproduce the values of the moments in Table~\ref{latticemoments}; and the positive value of the first moment indicates that, on average, the $s$-quark carries more of the bound-state's momentum than the $\bar u$-quark. In addition, the PDAs predict the quantities listed in Table~\ref{predictions}.

The positive value of $\langle (x-\bar x) \rangle_{su}$ is responsible for the shift in position, relative to the peak in the pion's PDA, of the maximum in $\varphi_{su}(x)$; viz., from $x=0.5$ to $x=0.55$, which is apparent in Fig.\,\ref{figNewLatticeK}.  This 10\% increase is a measure of nonperturbative SU$(3)$-flavour-symmetry breaking.  It is comparable with the 15\% shift in the peak of the kaon's valence $s$-quark PDF, $s_v^K(x)$, relative to $u_v^K(x)$ \cite{Nguyen:2011jy}.  By way of context, it is notable that the ratio of $s$-to-$u$ current-quark masses is approximately $28$ \cite{Beringer:1900zz}, whereas the ratio of nonperturbatively generated Euclidean constituent-quark masses is typically \cite{Chen:2012qr} $1.5$ and the ratio of leptonic decay constants $f_K/f_\pi \approx 1.2$ \cite{Beringer:1900zz}, both of which latter quantities are equivalent order parameters for dynamical chiral symmetry breaking (DCSB).  It is therefore apparent that the flavour-dependence of DCSB rather than explicit chiral symmetry breaking is measured by the shift in peak location.

Using the information presented above one can also report a lattice-QCD-based estimate for the ratio of kaon-to-pion elastic electromagnetic form factors at $Q^2=4\,$GeV$^2$ via Eqs.\,\eqref{pionUV}, \eqref{wphi}.  Let us first, however, provide some background.  Owing to charge conservation, $F_K(Q^2=0)/F_\pi(Q^2=0)=1$; and in the conformal limit, $F_K/F_\pi=f_K^2/f_\pi^2 = 1.50$.  Moreover, given that $r_\pi/r_K>1$, we anticipate that the ratio $F_K(Q^2)/F_\pi(Q^2)$ grows monotonically toward its conformal limit because anything else would indicate the presence of a new, dynamically generated mass-scale.  This expectation is supported by DSE form factor predictions \cite{Maris:2000sk}, which produce $F_K(\zeta_2^2)/F_\pi(\zeta_2^2)=1.13$.  Now, using Eqs.\,\eqref{pionUV}, \eqref{wphi} and the results in Table~\ref{predictions}, one has
\begin{equation}
\begin{array}{l|cc}
        & 16^3\times 32 & 24^3\times 64 \\\hline
\rule{0em}{3ex}
F_K(\zeta_2^2)/F_\pi(\zeta_2^2) & 1.21^{+1.22}_{-0.62} & 0.74^{+1.20}_{-0.51}
\end{array}.
\label{FKonFpi}
\end{equation}
The central value of this ratio obtained from the $16^3\times 32$ lattice is consistent with expectations and the DSE prediction; but the large errors on $\mathpzc{w}_P$; i.e., the $(1/x)$-moment, diminish the significance of this outcome.  Regarding the result obtained from the $24^3\times 64$ lattice, the central value suggests that this larger lattice produces a pion PDA which is too broad, consistent with the discussion in the penultimate paragraph of Sec.\,3.  However, given the even larger errors in this case, little can safely be concluded.

\smallskip

\noindent\textbf{5.$\;$ERBL evolution}.
As noted above, with decreasing $\tau=1/\zeta$, all meson PDAs shift pointwise toward $\varphi^{\rm asy}$ in Eq.\,\eqref{phiasy}.  This evolution was canvassed elsewhere for the symmetric pion PDA \cite{Cloet:2013jya,Cloet:2013tta}.  Herein, it is therefore interesting to elucidate the effect of evolution on the skewed kaon distribution associated with the moments produced by lattice-QCD.

The solid curve (labelled ``E'') in Fig.\,\ref{figEvolvedK} is the $24^3\times 64$-lattice kaon PDA, defined by the central values of $\alpha,\beta$ in Eq.\,\eqref{phiK24}, evolved to $\tau_{10}=1/\zeta_{10}$, $\zeta_{10}=10\,$GeV, using the leading-order ERBL equations.  The evolved distribution is described by
\begin{equation}
24^3 \times 64_{(\tau_2 \to \tau_{10})}{:}  \quad
\alpha_{su}=0.62^{+0.15}_{-0.13},\;
\beta_{su}=0.53^{+0.14}_{-0.12}, \quad
\end{equation}
and has a central-value peak-location shifted just 2.4\% closer to $x=\frac{1}{2}$.  It is apparent in Figure~\ref{figEvolvedK} that PDA evolution is slow.

\begin{figure}[t]
\centerline{\includegraphics[width=0.9\linewidth]{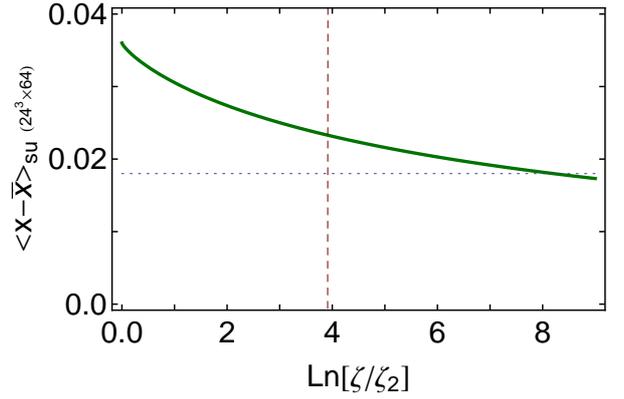}}
\caption{\emph{Solid curve} -- Leading-order evolution of $\langle x - \bar x\rangle_{su}$ with scale, $\zeta$, computed from the PDA defined by the central values in Eq.\,\eqref{phiK24}.  The vertical \emph{dashed line} marks $\zeta=\zeta_{100}$.  The horizontal \emph{dotted line} marks 50\% of this moment's $\zeta=\zeta_2$ value.  It is not reached until the energy scale $\zeta={\rm e}^{8.2} \zeta_2 = 7.3\,$TeV.
\label{figEvolvedKmom1}}
\end{figure}

The slow pace of evolution can be quantified as follows.  Consider the moment $\langle x - \bar x\rangle_{su}$, which measures the average excess of momentum carried by the valence $s$-quark in the meson.  As indicated above, this moment is a measure of the magnitude and flavour-dependence of DCSB.  It is zero in the conformal limit.  The $\zeta$-evolution of $\langle x - \bar x\rangle_{su}$ is depicted in Fig.\,\ref{figEvolvedKmom1}.  Plainly, the $s$-quark momentum-excess remains more than 50\% of its $\zeta_2$-value until energy scales exceeding those generated at the LHC.  Hence, consistent with similar analyses of $\varphi_{du}$, nonperturbative phenomena govern the pointwise behavior of $\varphi_{su}$ at all energy scales that are currently conceivable in connection with terrestrial facilities.  (Higher-order evolution \cite{Mikhailov:1984ii,Mueller:1998fv} does not materially affect these results or conclusions.)

Finally, we return to the ratio in Eq.\,\eqref{FKonFpi} and consider the impact of ERBL evolution.  Working with the $16^3\times 32$ lattice, which produces the more reasonable value, the ratio in Eq.\,\eqref{FKonFpi} becomes
\begin{equation}
\label{FKratio10}
F_K(\zeta_{10}^2)/F_\pi(\zeta_{10}^2) = 1.29^{+0.61}_{-0.42}\,.
\end{equation}
Subtracting the unit $Q^2=0$ value, guaranteed by charge conservation, Eq.\,\eqref{FKratio10} describes a 38\% increase in the central value of the ratio.  Notably, the error band has narrowed along with the distributions.  At $\zeta_{100}=100\,$GeV; i.e., $Q^2=10\,000\,$GeV$^2$, the central value of the ratio is $1.34$, which is a further increase of 17\%.  However, one still remains at only 68\% of the pertinent conformal limit value.

\smallskip

\noindent\textbf{6.$\;$Epilogue}.
Light-front parton distribution amplitudes (PDAs) have numerous applications in the analysis of hard exclusive processes in the Standard Model but predictive power is lacking unless they can be calculated.  Many nonperturbative methods for the estimation of nonperturbative matrix elements in QCD produce moments of the PDAs, instead of the pointwise behaviour directly.  Therefore, in order to make progress, one needs an effective means by which to reconstruct the PDA from its moments.

The method introduced in Refs.\,\cite{Chang:2013pq,Cloet:2013tta,Chang:2013epa} enables one to obtain a pointwise accurate approximation to meson PDAs from limited information.  We employed it to extract the PDAs of light-quark pseudoscalar and vector mesons from the restricted number of moments made available by numerical simulations of lattice-regularised QCD.  Our analysis shows that, at all energy scales currently accessible to terrestrial experiments, the PDAs are concave functions whose dilation and asymmetry, when the latter is present, express the strength of dynamical chiral symmetry breaking.

Notably, within errors, the lattice moments indicate that when constituted from the same valence quarks, the PDAs of pseudoscalar and vector mesons are identical.  Some studies in continuum QCD support an approximate equality between these amplitudes, however, there is significant disagreement between methods and models.  In addition, the lattice moments appear precise.  However, our analysis showed this appearance to be misleading because the errors on the moments admit an error band on a given PDA which is effectively large.  Moreover, especially for the pion, the lattice moments exhibit material dependence on lattice volume.  It is plain, therefore, that improvements must be made in both continuum- and lattice-QCD before we arrive at an accurate, unified picture of the light-front structure of pseudoscalar and vector mesons constituted from light-quarks.


\smallskip

\noindent\textbf{Acknowledgments}.
We are grateful for insightful comments from G.~Gao, Y.-x.~Liu and P.\,C.~Tandy; and for the enthusiasm and hospitality of students, especially Z.-f.~Cui, in the Physics Department at Nanjing University where this work was conceived and partially completed.
Work supported by:
University of Adelaide and Australian Research Council through grant no.~FL0992247;
For\-schungs\-zentrum J\"ulich GmbH;
Department of Energy, Office of Nuclear Physics, contract no.~DE-AC02-06CH11357;
the National Natural Science Foundation of China (Grant nos.\ 11275097, 10935001 and 11075075); and the Research Fund for the Doctoral Program of Higher Education (Grant no.\ 2012009111002).

\smallskip



\end{document}